\documentclass[fleqn,usenatbib]{mnras}

\usepackage{newtxtext,newtxmath}

\usepackage[T1]{fontenc}
\usepackage{ae,aecompl}

\usepackage{graphicx}	
\usepackage{amsmath}	
\usepackage{hyperref}
\usepackage{upgreek}
\usepackage{gensymb}
\usepackage{subfig,caption}
\usepackage[section]{placeins}
\newcommand{\mydelta}{$\Delta_{\rm mdl - obs}$}
\newcommand{\darkeningslopes}{34$\pm$4, 46$\pm$2 and 53$\pm$2 milli-mag$_{\rm SQM}$\,arcsec$^{-2}$\,yr$^{-1}$ for Stockholm, Potsdam-Babelsberg and Vienna}
\newcommand{\magsqm}{mag$_{\rm SQM}$\,arcsec$^{-2}$}

\title[Correct for darkening of SQM data using twilight]{Correcting Sky Quality Meter measurements for aging effects using twilight as calibrator}

\author[Puschnig, N\"aslund, Schwope, Wallner]{
Johannes Puschnig,$^{1}$\thanks{E-mail: johannes@sonnensystem.at}
Magnus N\"aslund,$^{2}$
Axel Schwope,$^{3}$
Stefan Wallner$^{4}$
\\
$^{1}$Universit\"at Bonn, Argelander-Institut f\"ur Astronomie, Auf dem H\"ugel 71, D-53121 Bonn, Germany\\
$^{2}$Department of Astronomy, Stockholm University, AlbaNova University Centre, SE-10691 Stockholm, Sweden \\
$^{3}$Leibniz-Institut f\"ur Astrophysik Potsdam (AIP), An der Sternwarte 16, 14482 Potsdam, Germany \\
$^{4}$Universit\"at Wien, Institut f\"ur Astrophysik, T\"urkenschanzstra{\ss}e 17, A-1180 Wien, Austria\\
}

\date{Submitted to MNRAS, Dec 7, 2020}

\pubyear{2020}

\begin{document}
\label{firstpage}
\pagerange{\pageref{firstpage}--\pageref{lastpage}}
\maketitle

\begin{abstract}
In the last decade numerous Sky Quality Meters (SQMs) were installed throughout the globe,
aiming to assess the temporal change of the night sky brightness (NSB), and thus the change in light pollution.
However, it has become clear that SQM readings may be affected by aging effects such as degradation of the sensor
sensitivity and/or loss of transmissivity of optical components (filter, housing window).
To date, the magnitude of the darkening has not been assessed in a systematic way. We report
for the first time on the quantification of the SQM aging effect and
describe the applied method.
We combine long-term SQM measurements obtained
between 2011 and 2019 in Potsdam-Babelsberg (23 km to the southwest of the center of Berlin),
Vienna and Stockholm with
a readily available empirical twilight model, which serves as calibrator.
Twilight SQM observations, calibrated for changing sun altitudes,
reveal a linear degradation of the measurement systems (SQM + housing window) with the following
slopes: \darkeningslopes.
With the highest slope found in Vienna (latitude $\sim$48$^\circ$) and the lowest
one found in Stockholm (latitude $\sim$59$^\circ$), we find an indication for
the dependence of the trend on solar irradiance (which is a function of
geographic latitude).
\end{abstract}

\begin{keywords}
light pollution -- atmospheric effects -- methods: data analysis
\end{keywords}



\section{Introduction}
It was shown by numerous studies that the ever increasing amount of \textit{artificial light at night} (ALAN) has far reaching
consequences, not only for astronomy, energy consumption and carbon footprint, but also for animal
and human health \citep{Chepesiuk2009,Haim2013,Cho2015,Garcia-Saenz2018},
ecosystems \citep{Longcore2004} and biodiversity \citep{Hoelker2010}.
Because ALAN impacts life on Earth in such a drastic way, and on a global scale, it was proposed
to monitor light pollution in a similar manner as other pollutants. Although no uniform measuring standard
has been implemented to date, in recent years, a number of individuals, observatories and organisations
have started to monitor the night sky brightness (NSB) using different methods and devices \citep{Haenel2018},
of which the so called \textit{Sky Quality Meter} (SQM), is probably the most widely used one.

SQM networks of larger scale have been established, for example, in Upper Austria\footnote{\url{https://www.land-oberoesterreich.gv.at/159659.htm}},
where a total of 23 SQMs are operational since end of 2015 \citep{Posch2018,Puschnig2020},
with most of the SQMs mounted at weather stations.
Similarly, starting end of 2009, in Galicia (Spain) a number of 20 SQMs were installed at weather
stations of the official Galician meteorological agency\footnote{\url{http://www.meteogalicia.gal/Caire/brillodoceo.action}}
in cooperation with the University of Santiago de Compostela \citep{Bara2019b}. Being part of a meteorological monitoring system
makes these two networks to showcases for future monitoring campaigns around the globe.
However, other SQM networks of similar scale exist:
The Spanish Light Pollution Research collaboration\footnote{\url{http://guaix.fis.ucm.es/splpr/SQM-REECL}} runs 18 devices,
the `NachtMeetnet' network \citep{Schmidt2020} counts 15 SQMs, the University of Hong Kong\footnote{\url{http://nightsky.physics.hku.hk/}}
owns 19 stations, and seven SQMs are permanently installed in Italy's Veneto region \citep{Bertolo2019}.
Many other small scale networks or single SQMs are found around the globe. \cite{Kyba2015} have
compiled many of these heterogeneous data sets and analyzed it in a consistent way to quantify the properties of skyglow.

At present, scientific interpretation in terms of long-term trend analysis of the large amount of readily available SQM data,
is hindered by the fact that the magnitude of degradation of SQM measurements
(due to change in sensor sensitivity, filters and/or housing window) on operation timescales of several years
is unknown. It is thus a timely matter to understand how SQMs loose the initial calibration
when operated over multiple years. In this paper, we quantify the aging effect for the first time,
using long-term SQM measurements, re-calibrated during post-processing
with twilight models.


\section{Measurements and Data}
We are operating SQMs in Potsdam-Babelsberg (23 km to the southwest of the center of Berlin),
Vienna and Stockholm since 2011, 2012 and 2015 respectively, aiming to
monitor light pollution over long time scales.
The exact geographical coordinates of the measurement stations are given in Table \ref{tab:sqmcoords}.
As the atmospheric composition (aerosols, ozone, particulate matter) plays a major role
in interpreting NSB measurements, we further use meteorological parameters obtained 
through the Copernicus Climate Change Service (C3S) information 2020 and
the Copernicus Atmosphere Monitoring Service (CAMS) Information 2020 \citep{CAMS}, in order
to monitor any systematic changes of the atmospheric composition within the
period of time under consideration.

\subsection{SQM measurements}
A first analysis of the
SQM data obtained in Potsdam-Babelsberg and Vienna was previously presented in \citep{Puschnig2014b} and \citep{Puschnig2014a},
and some of the Stockholm data was already discussed in \cite[Section 4.3 therein]{Posch2018}.

The exact model designation is SQM-LE, indicating that the devices are equipped with a front lens (L)
and ethernet (E) connector. The SQMs are permanently installed and point towards the zenith. Given the Gaussian-like
angular response as described in \cite{Cinzano2007} the reported radiances in units of \magsqm\ are representative
for the average zenithal sky brightness found in a circum-zenithal region with a radius of 10 degrees.
Our SQMs are controlled via home-brewed \texttt{Perl} scripts, producing measurements at frequencies of
approximately 0.14 (IFA, STO) and 0.5 (BA1) Hz.
Further technical details are found in \cite{Cinzano2005} and \cite{Bara2019b}.

\begin{table}
\centering
\caption{Locations, station codes, device serial numbers (SN) and geographical coordinates (lat, lon) of the SQMs}
\label{tab:sqmcoords}
\begin{tabular}{l l c r r}
Country, City, District           & Code  & SN    & Lat           & Lon        \\
\hline \hline
SE, Stockholm, \"Ostermalm    & STO   & 2785  & 59$^\circ$21'12"N  & 18$^\circ$3'28"E \\
DE, Potsdam, Babelsberg       & BA1   & --    & 52$^\circ$22'48"N  & 13$^\circ$6'22"E \\
AT, Vienna, W\"ahring        & IFA   & 1898  & 48$^\circ$13'54"N  & 16$^\circ$20'3"E \\
\end{tabular}
\end{table}

\subsection{Archival Meteorological Data}
Large-scale atmospheric and meteorological parameters were obtained through \texttt{MARS}, the
Meteorological Archival and Retrieval System, which provides users with
meteorological forecast and analysis results
from the European Centre for Medium-Range Weather Forecasts (ECMWF) in
\texttt{GRIB} format.
In particular, we retrieved parameters
from ECMWF's re-analysis product 
`ERA5' and atmospheric composition (aerosol optical depths, particulate matter)
from `CAMS Near-real-time' data.
ERA5 provides data with an hourly validity. However, particulate matter and optical depths,
are available only through forecast models with a validity given in 3-hour steps.
The spatial resolution of the data grid is 30 and 80km for ERA5 and CAMS Near-real-time respectively.
Furthermore, some parameters (e.g. particulate matter) are only available starting from
2015. Relevant parameters for the three cities of Vienna, Potsdam and Stockholm are shown in Figures
\ref{fig:meteo1}--\ref{fig:meteo3}.

\section{Models and Methods}\label{sec:models_methods}
In order to reveal the existence of darkening of SQM measurements over time,
we use empirical twilight models previously published by \cite{Patat2006} as calibration source.
In particular, the temporal change of the difference between predicted NSB (mdl) and the observed one (obs), \mydelta,
will be used as a proxy of the aging effect.
The models are based on a statistically significant sample
of thousands of \textit{UBVRI} (Johnson-Cousins system) twilight sky flats obtained during
several years on mount Paranal (Chile) as part of an instrument calibration procedure.
Hence, they are valid only for a limited range of solar zenith distances (z),
i.e. 95--105$^\circ$. The functional (polynomial) form for the zenithal NSB during twilight
is shown in equation \ref{equ:patatmodel}. The coefficients for the \textit{BVR} filters are found in Table \ref{tab:patatmodel_coeff}.
\begin{equation}\label{equ:patatmodel}
    NSB = a_0 + a_1 * (z-95) + a_2 * (z-95)^2
\end{equation}

\begin{table}
\centering
\caption{Coefficients of the empirical polynomial twilight model of Patat et al. (2006) for the Johnson-Cousins $B$, $V$ and $R$ filter.}
\label{tab:patatmodel_coeff}
\begin{tabular}{l r r r}
filter & a$_0$ & a$_1$ & $a_2$ \\
\hline \hline
B & 11.84 & 1.411 & -0.041 \\
V & 11.84 & 1.518 & -0.057 \\
R & 11.40 & 1.567 & -0.064 \\
\end{tabular}
\end{table}

Given the fact that the calculation of atmospheric diffuse flux
is a rather complicated problem that requires a detailed treatment of multiple
scatterings \citep{Kocifaj2019}
and knowledge of atmospheric composition such as aerosols, ozone and particulate matter \citep{Joseph1991,Sciezor2014,Sciezor2020b}.
it is expected that the residuals of \mydelta\ show discrepancies.
However, the \textit{average} magnitude of these discrepancies should be independent of time,
as long as the \textit{average} atmospheric composition has not changed significantly.
In that case, and as long as a large number of independent observations were carried out,
one would expect that the residual of \mydelta, measured over multiple years,
is given by a constant, plus scatter with a roughly constant magnitude, caused by complex atmospheric physics.
We may thus formulate the null hypothesis that any temporal trend (slope) in the
residual (twilight model minus SQM measurements) is caused by instrumental effects
(i.e. the SQM aging effect), as long as the \textit{average} composition of the atmosphere has not changed
significantly (towards lower abundances) within the same period of time.
We would further expect that the degree of darkening (aging effect) is 
a function of solar surface radiance. Given the range in geographic latitudes of our
measurement stations it is thus expected that the effect of darkening is weakest in Stockholm
and strongest in Vienna.

In addition to the previously described scatter of \mydelta, clouds have
a strong impact on NSB \citep{Puschnig2014b, Jechow2017, Jechow2019}. We thus aim to
select only clear-sky twilight measurements. To do so, we slice our observations
into 5-minute data chunks and evaluate the standard deviation after subtraction of a linear fit.
The latter is used as a proxy for cloud coverage. Note that it was previously shown by
\cite{Cavazzani2020} that the standard deviation of SQM measurements
may serve as clear sky indicator.
For the clear-sky case, the following
assumptions are made: the maximum deviation is lower than 0.06\,\magsqm\ and
the standard deviation is lower than 0.02*sqrt(sampling$_{nominal}$/sampling), with the
nominal sampling frequency being 0.14\,Hz (IFA, STO).
In order to avoid any influence ALAN might have,
we only consider twilight observations with sun altitudes between -6 and -7$^\circ$,
i.e. the bright end of the models with NSBs of approximately 13--14.5\,\magsqm.
Given typical clear-sky NSBs between 18--20\,\magsqm\ observed in the three cities,
our twilight measurements fall within brightness levels that are at least
a factor of 25 higher than those used for the assessment of light pollution.
Any contribution of ALAN to our measurements is thus negligible. 
Analogue, we consider only data with moon altitudes lower than -5$^\circ$.
The final set of data points (\mydelta) are thus the 5-minute averages of
the difference between the model and the SQM observations, under clear sky.

\section{Results}
\begin{figure}
    \subfloat{\includegraphics[width=0.8\columnwidth]{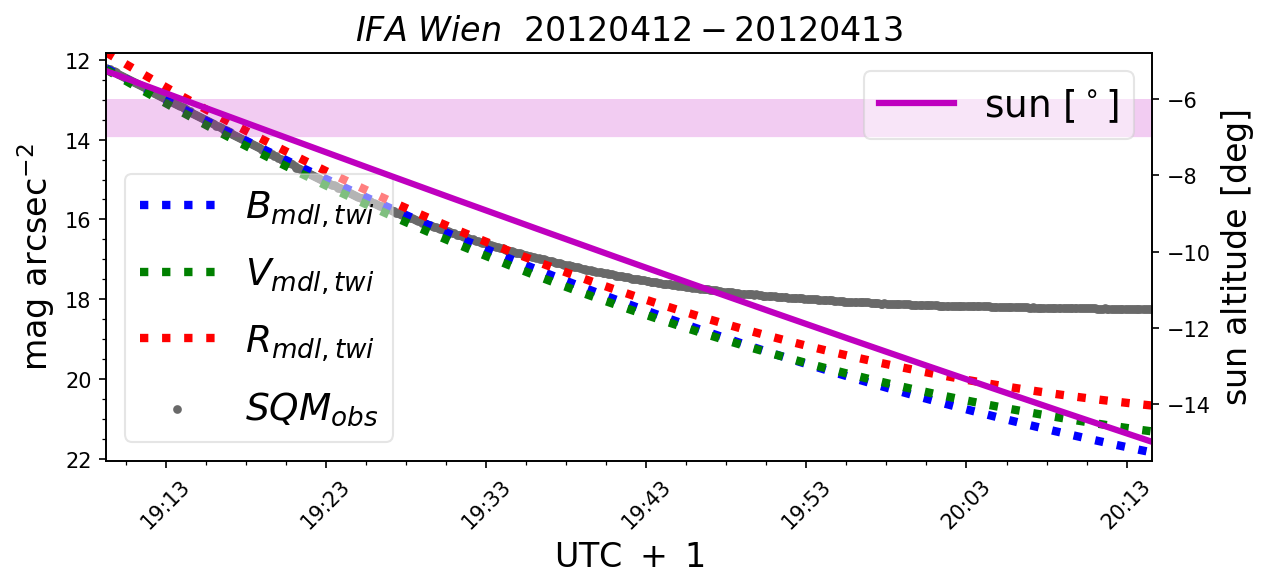}}\hspace*{\fill}
    
    \subfloat{\includegraphics[width=0.8\columnwidth]{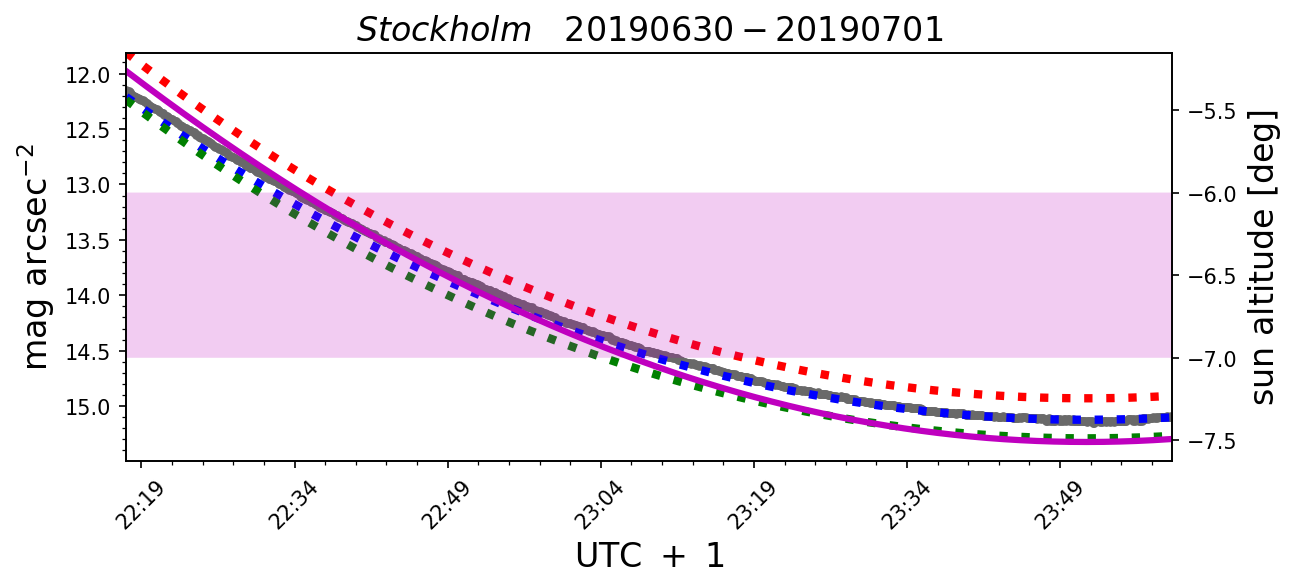}}\hspace*{\fill}
    \caption[SQM measurements Vienna and Stockholm]{Exemplary scotographs of SQM measurements (\textit{gray curve}) obtained during the course of dusk under clear sky
        in Vienna on April 12, 2012 (\textit{top panel}) and in Stockholm on June 30, 2019 (\textit{bottom panel}).
        Twilight \textit{BVR} models are shown in \textit{blue}, \textit{green} and \textit{red} respectively. The \textit{magenta}
        solid curve indicates the solar altitude and
        the highlighted area is the sun altitudes interval [-6$^\circ$,-7$^\circ$] that was used for calibration.}
    \label{fig:twilight}
\end{figure}

%
%
%
A qualitative comparison between the twilight models as described in Section \ref{sec:models_methods}
and two selected scotographs obtained on April 12, 2012 in Vienna and on June 30, 2019 in Stockholm,
demonstrates that the polynomial models are indeed an eligible approximation for the true decline of zenithal NSB
as observed with the SQM (see Figure \ref{fig:twilight}), at least as long as ALAN's contribution to NSB is negligible.
The figure also shows that the functional form of
the modeled gradual decline is very similar in all three filters, in particular
in the regime of interest for this work, i.e. when the sun is found between -6 and -7$^\circ$.
However, the absolute difference between models and data is smallest in ($B-SQM$). Therefore,
we decided to use the $B$-band models as a reference.
The top panel in Figure \ref{fig:twilight} further reveals that, in Vienna under clear sky, light pollution starts
to dominate the observed NSB once the Sun has declined more than approximately 10 degrees below
the horizon. Note that at this point the observed brightness levels are already 2.5--3.5\,\magsqm\ higher
(darker) than during the range that is used for the calibration of the SQM measurements during post-processing.

\begin{figure}
    \subfloat{\includegraphics[width=0.8\columnwidth]{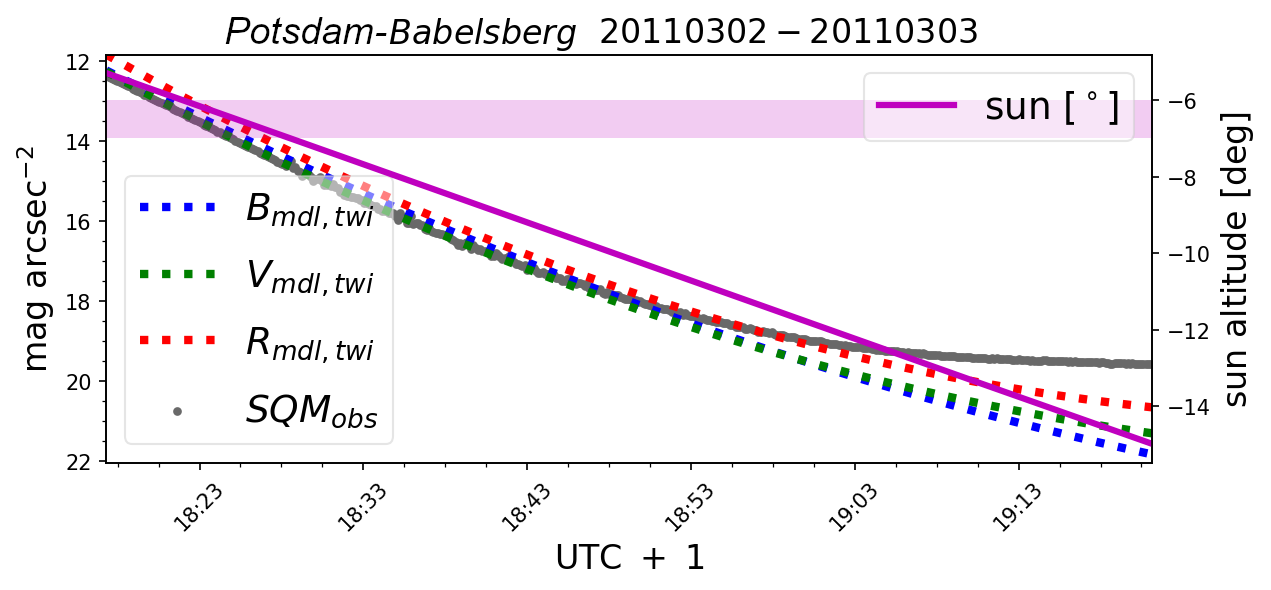}}\hspace*{\fill}
    
    \subfloat{\includegraphics[width=0.8\columnwidth]{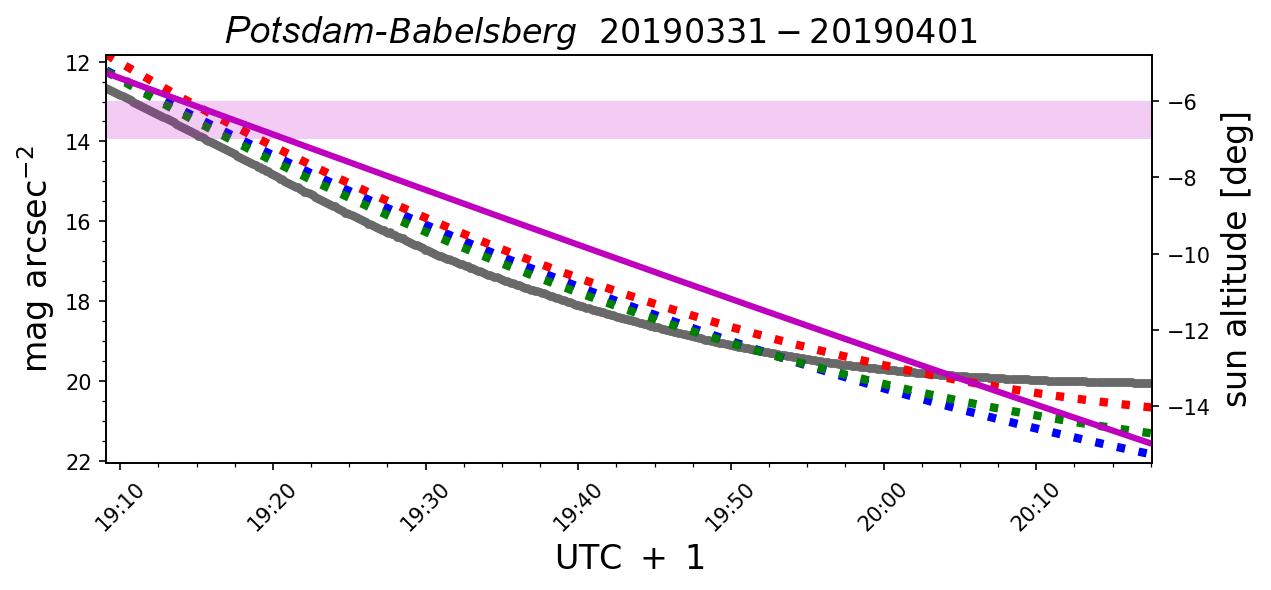}}\hspace*{\fill}
    \caption[SQM measurements Potsdam 2011, 2019]{Scotographs of SQM measurements (\textit{gray curve}) obtained during the course of dusk under clear sky
        in Potsdam on March 2, 2011 (\textit{top panel}) and March 31, 2019 (\textit{bottom panel}). The comparison qualitatively shows the darkening effect
        after 9 years of operation. For more details, see caption of Figure \ref{fig:twilight}.
        }
    \label{fig:twilight_berlin}
\end{figure}

\begin{figure*}
        \begin{center}
        \includegraphics[width=0.8\textwidth]{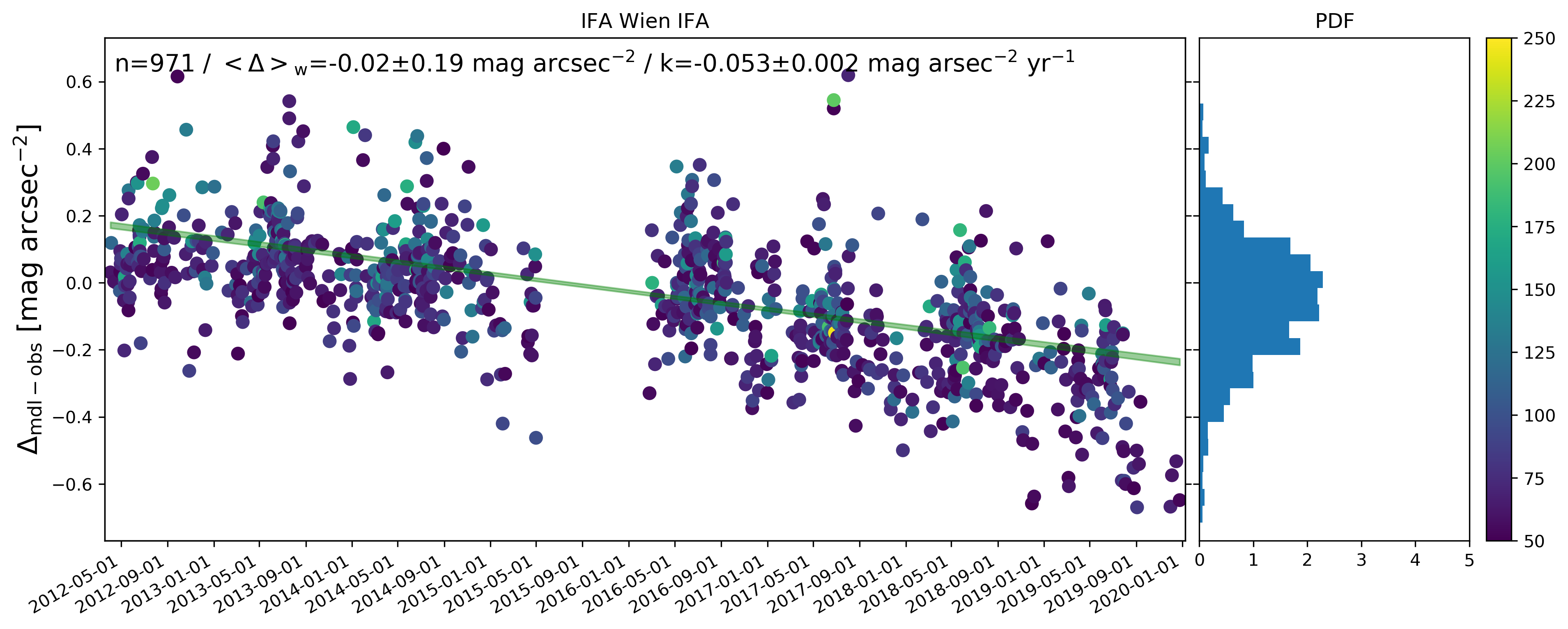}
        \caption[SQM Aging Function Vienna]{SQM aging function for Vienna (latitude $\sim$48$^\circ$), derived from calibrating SQM measurements (obtained between 2012 and 2019)
        with twilight models.
        Time (x-axis) vs. difference of modeled $B$-filter radiance minus observed SQM radiance (y-axis). Each point is the 5-minute average value obtained under \textit{clear sky}
        and within a narrow interval of sun altitudes [-6$^\circ$,-7$^\circ$]. The color represents
        the inverse variance of the underlying SQM data (5-minute time intervals). For example, a value higher than 100 means that the standard deviation (of the model-observation difference)
        within the time interval was lower than 0.1\,\magsqm. The green shaded region is the result
        from a weighted linear fit through the data including the +/-1-sigma error range, using the inverse variance as weights.}
        \label{fig:Vienna_aging}
        \end{center}
\end{figure*}

\begin{figure*}
        \begin{center}
        \includegraphics[width=0.8\textwidth]{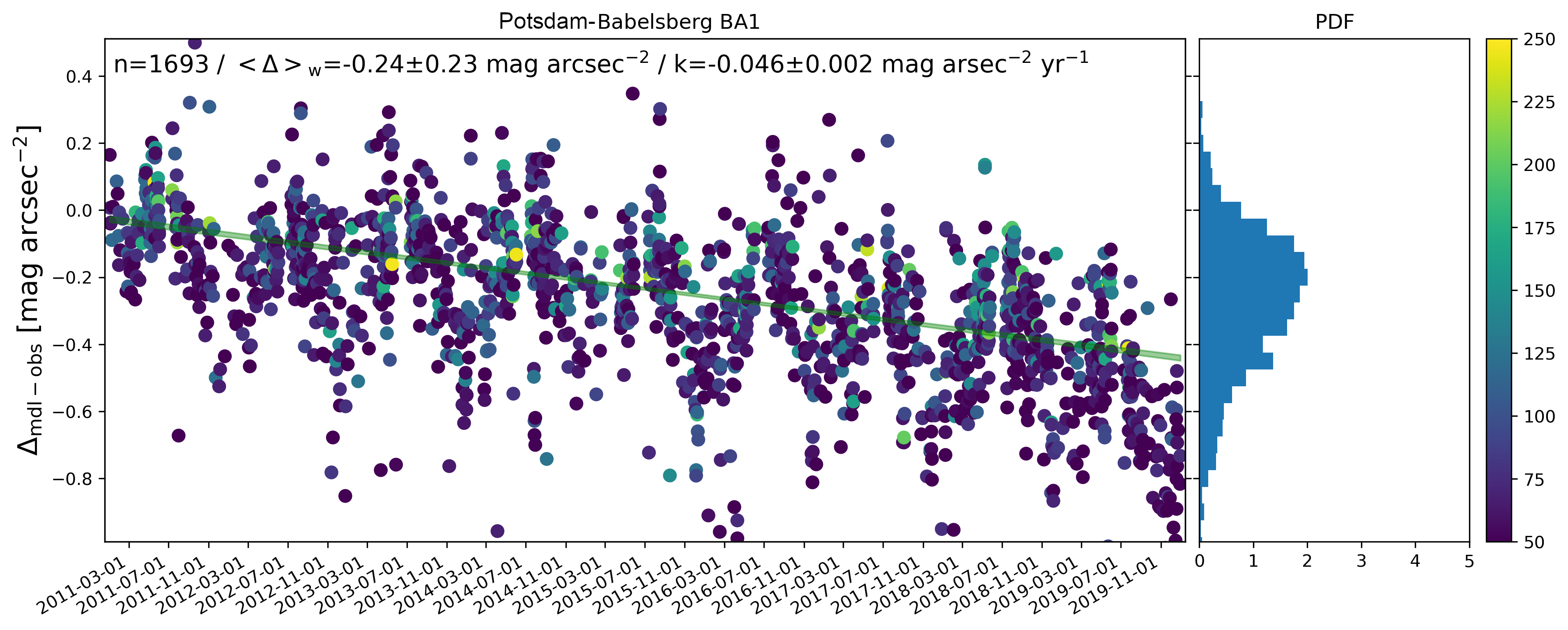}
        \caption[SQM Aging Function Potsdam]{SQM aging function for Potsdam (latitude $\sim$52$^\circ$), derived from calibrating SQM measurements (obtained between 2011 and 2019)
        with twilight models.
        See caption of Figure \ref{fig:Vienna_aging} for more details.}
        \label{fig:Berlin_aging}
        \end{center}
\end{figure*}

\begin{figure*}
        \begin{center}
        \includegraphics[width=0.8\textwidth]{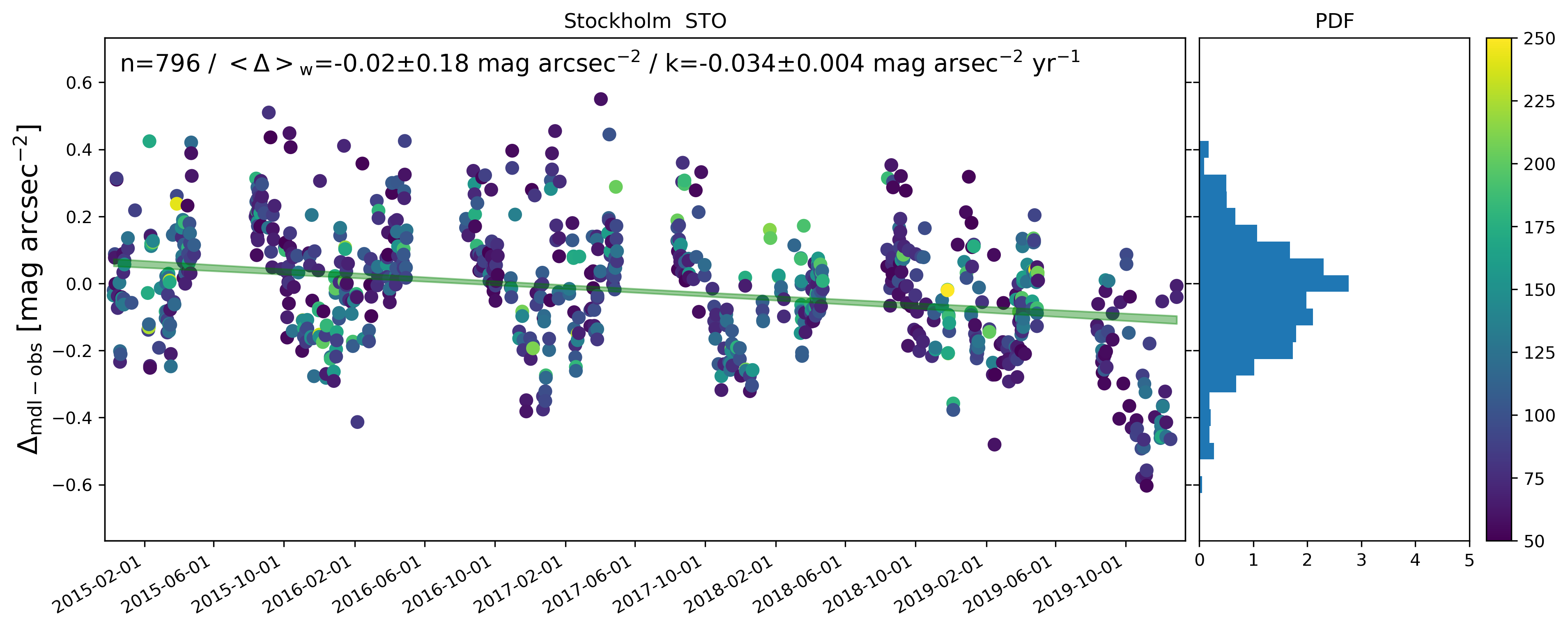}
        \caption[SQM aging Function Stockholm]{SQM aging function for Stockholm (latitude $\sim$59$^\circ$), derived from calibrating SQM measurements (obtained between 2015 and 2019)
        with twilight models.
        See caption of Figure \ref{fig:Vienna_aging} for more details.}
        \label{fig:Stockholm_aging}
        \end{center}
\end{figure*}

Resembling previous results of \cite{Puschnig2014b}, the overall level of light pollution (deduced
from zenithal NSB) is much lower for the SQM station in Potsdam. The scotograph in the top panel of Figure \ref{fig:twilight_berlin}
demonstrates that ALAN starts to dominate from sun altitudes of approximately -12 degrees only.
When browsing through the available scotographs\footnote{All data is available for download and examination via \url{https://astro.univie.ac.at/en/about-us/light-pollution/}},
one may already qualitatively assess the effect of darkening, in particular for the 9-year long data series of Potsdam (see Figure \ref{fig:twilight_berlin}).
We stress that this is only a showcase-comparison, because the atmospheric composition has a strong impact
on NSB. And given the fact that the atmospheric composition may change on a daily and seasonal basis, it is
expected that the aging effect itself is buried in a large scatter. However, 
the null hypothesis is that the average of \mydelta\ should remain constant
over multiple years, as long as the atmospheric composition did not systematically change.
On the other hand, any time-dependent trend of \mydelta\ must
then be caused by instrumental effects, i.e. instrumental aging of the device due to permanent exposure to solar irradiance.
After filtering and slicing our SQM
measurements into clear-sky chunks with a length of 5 minutes
as explained in Section \ref{sec:models_methods}, a relatively strong trend
(see Figures \ref{fig:Vienna_aging}--\ref{fig:Stockholm_aging}) is revealed, suggesting that the
SQM measurements are indeed affected by average darkening levels of \darkeningslopes. Note that the linear regressions in
Figures \ref{fig:Vienna_aging}--\ref{fig:Stockholm_aging} were derived under usage of weights, with the latter
given by the variance of the data within the 5 minute long chunks. The fitting results are presented in
Table \ref{tab:fittingresults}. We stress that the shape of the probability density distributions of NSB
values in the right panels of Figures \ref{fig:Vienna_aging}--\ref{fig:Stockholm_aging}
are indicative for the successful removal of data affected by clouds. Because in the latter case, one would expect an
asymmetric distribution (log-normal with negative skewness) of NSB values (compare \citealp{Sciezor2020a}).

\begin{table}
\centering
\caption{SQM aging functions from weighted linear regression}
\label{tab:fittingresults}
\begin{tabular}{l r r}
station  & slope &                 intercept        \\
 code     & [\magsqm yr$^{-1}$]   & [\magsqm] \\
\hline \hline
STO   & -0.034 $\pm$ 0.004 & 68 $\pm$8 \\
BA1   & -0.046 $\pm$ 0.002 & 92 $\pm$3 \\
IFA   & -0.053 $\pm$ 0.002 & 106 $\pm$4 \\
\end{tabular}
\end{table}

\section{Discussion}
Beside the aging trend, Figures \ref{fig:Vienna_aging}--\ref{fig:Stockholm_aging} show
strong seasonal NSB variations, in particular for Stockholm. It seems that the amplitude of
the seasonal variation is a function of geographic latitude as well. A possible explanation
for that may be the increase of snow coverage and snow depth with latitude, leading to
stronger variations in surface albedo, that in turn have strong impact on the NSB measurements (compare
\citealp{Wallner2019}).

The Stockholm data also show another peculiarity, i.e. gaps in data during summer.
The reason is the stringent filtering for clear-skies using an upper limit standard
deviation (after subtraction of a linear fit from the SQM data)
within 5-minute intervals.
In Stockholm, for few days around summer solstice, the Sun's rate of decline
per unit time is far from being constant (within the interval of interest).
In this case, the polynomials of the \cite{Patat2006} twilight models appear to
systematically differ from the observations.

Finally, it is seen in Figures \ref{fig:Vienna_aging}--\ref{fig:Stockholm_aging}
that \textit{individual measurements} obtained under clear sky may differ by a large amount,
even up to 1\,\magsqm\ (or a factor of 2.5) on timescales of days or weeks. This scatter
is mainly caused by complex atmospheric physics (e.g. short-term changes of atmospheric composition).
It is thus important having obtained a statistically significant number of measurements
in order to perform an unbiased long-term trend study.

\section{Summary and Conclusion}
Aiming to study the darkening of SQM measurements with time,
we have compared readily available twilight models \citep{Patat2006} with
long-term (5--9 years) zenithal NSB measurements obtained during twilight with SQMs
at three different sites (Vienna, Potsdam-Babelsberg and Stockholm).
Using only measurements obtained at solar zenith angles between 96 and 97$^\circ$,
allows us to focus on brightness levels in which ALAN's contribution to the zenithal NSB
is negligible (likely true for most sites on Earth). Further slicing and filtering
the data into clear-sky chunks (using the standard deviation as a clear sky
indicator), reveals not only strong seasonal variations, but also an instrumental
darkening effect that is well described by a linear
fit with slopes of \darkeningslopes\ (see Figures \ref{fig:Vienna_aging}--\ref{fig:Stockholm_aging}).
Given the strong
influence of atmospheric composition on NSB, we have checked large-scale
atmospheric parameters (AOD, particulate matter, ozone) for existing trends that
might cause the observed darkening. As shown in Figures \ref{fig:meteo1}--\ref{fig:meteo3}
no significant trend was found that could possibly lead to the observed slopes.
However, we do notice a slight upward trend of particulate matter, apparent at
all three sites. This might have some impact on the derived slopes, due to an
increased fraction of (back-)scattering of light, and would even increase the
observed aging slopes.

\section*{Acknowledgements}
Part of this work has been generated using Copernicus Climate Change Service information 2020 and
Copernicus Atmosphere Monitoring Service Information 2020.




\bibliographystyle{mnras}
\bibliography{lp} 


\FloatBarrier
\appendix
\section{Additional Tables and Figures}
See next page.
\begin{figure*}
    \subfloat{\includegraphics[width=0.87\textwidth]{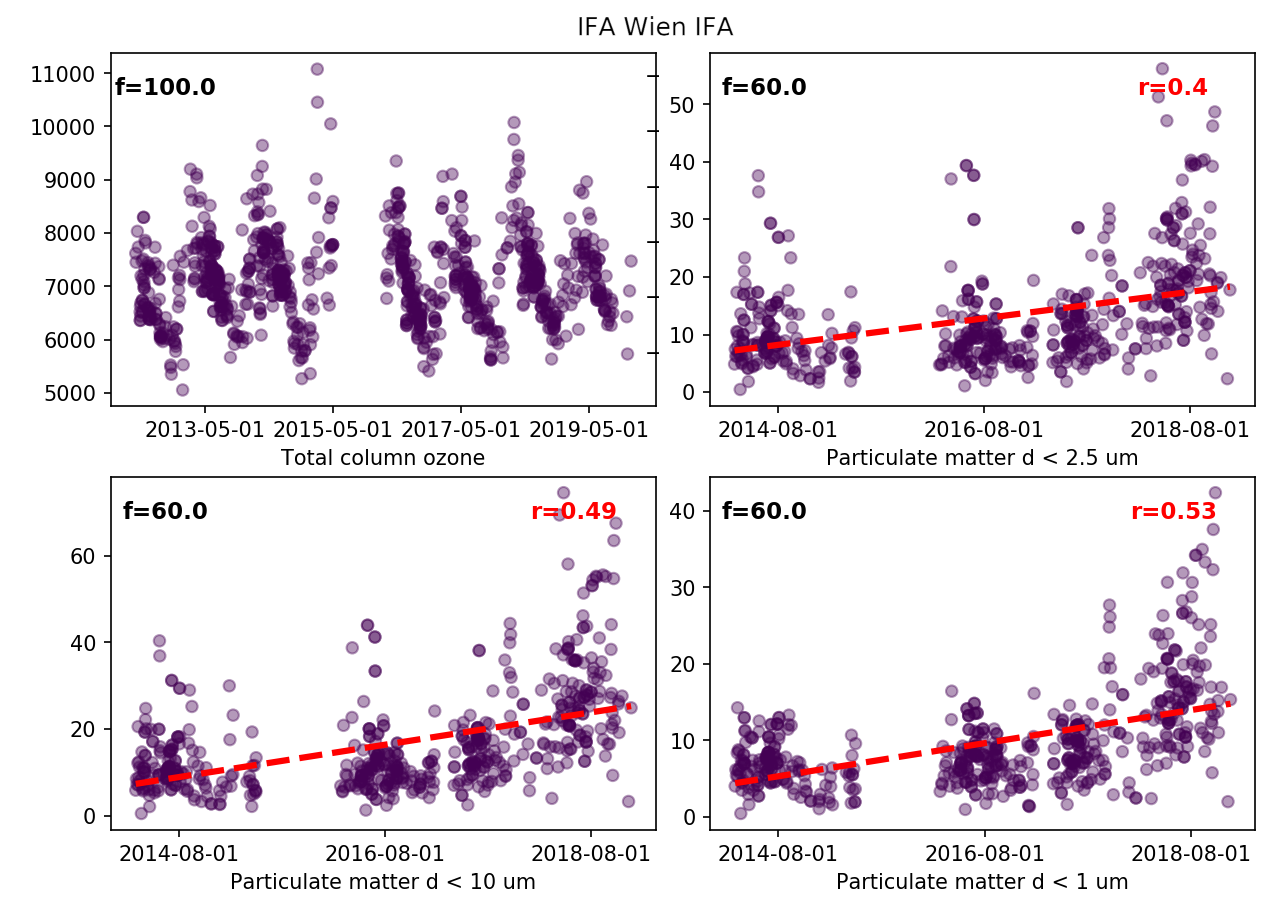}}
    
    \subfloat{\includegraphics[width=0.85\textwidth]{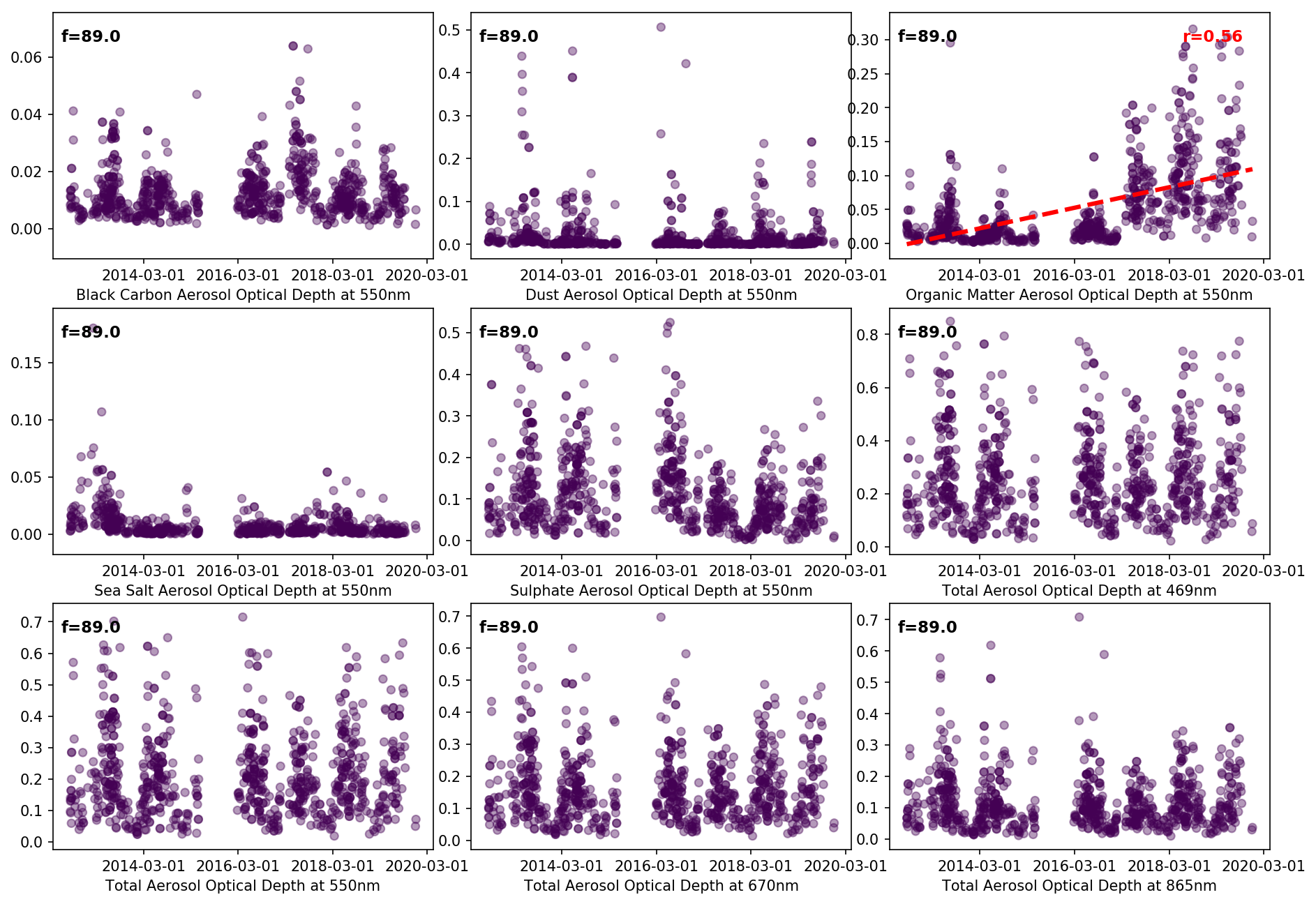}}
        
    \caption[Vienna Meteo]{Large-scale particulate matter (\textit{top panel}) and aerosol optical depth (\textit{bottom panel}) development for Vienna.
    The parameter `f' in the top left corner indicates what fraction of available SQM measurements could be matched with the meteorological data shown here.
    The y-axes units are $nm$ for molecular columns and $g\ cm^{-3}$ for particulate matter.}
    \label{fig:meteo1}
\end{figure*}

\begin{figure*}
    \subfloat{\includegraphics[width=0.87\textwidth]{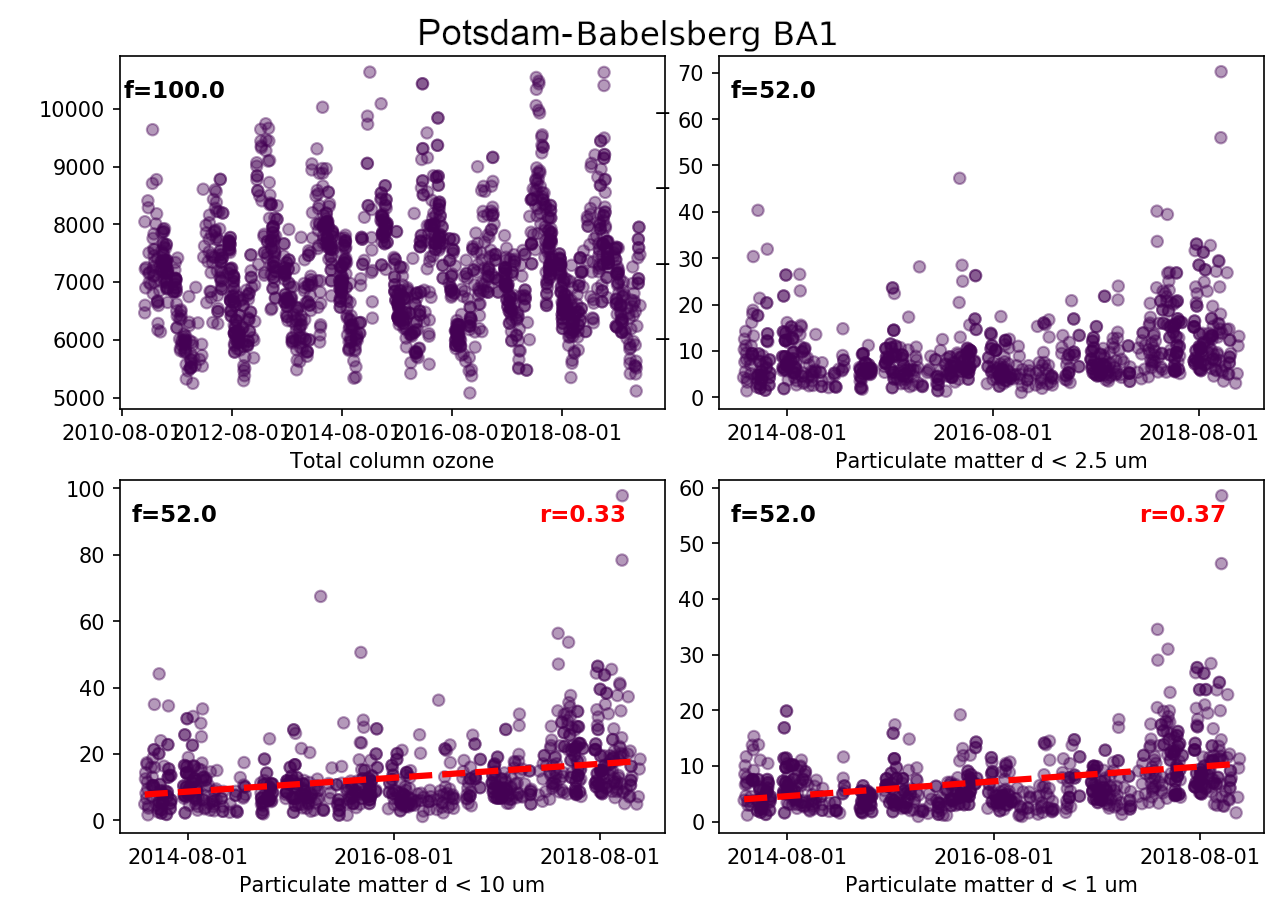}}
    
    \subfloat{\includegraphics[width=0.85\textwidth]{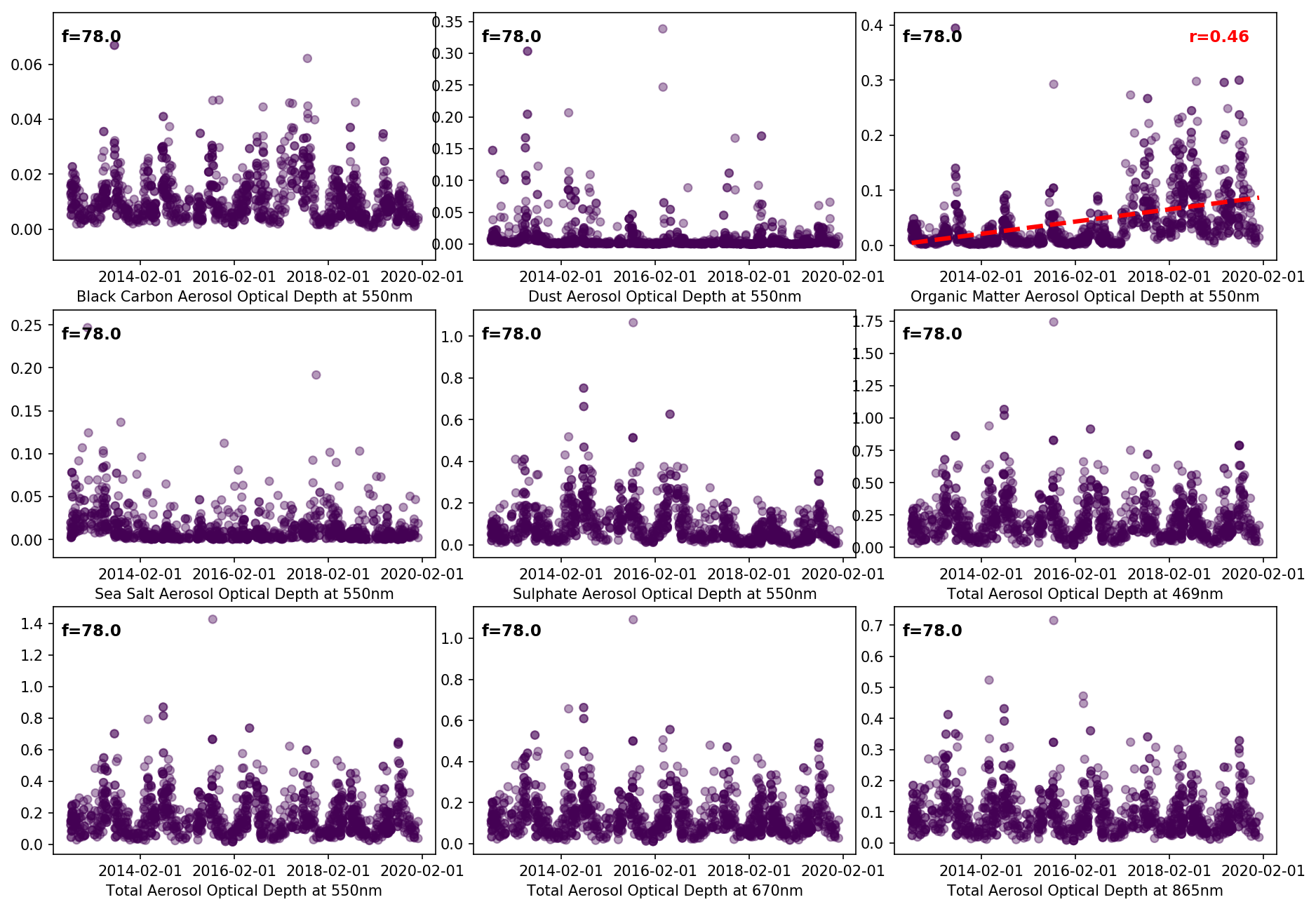}}
        
    \caption[Potsdam Meteo]{Large-scale particulate matter (\textit{top panel}) and aerosol optical depth (\textit{bottom panel}) development for Potsdam.
    The parameter `f' in the top left corner indicates what fraction of available SQM measurements could be matched with the meteorological data shown here.
    The y-axes units are $nm$ for molecular columns and $g\ cm^{-3}$ for particulate matter.}
    \label{fig:meteo2}
\end{figure*}

\begin{figure*}
    \subfloat{\includegraphics[width=0.87\textwidth]{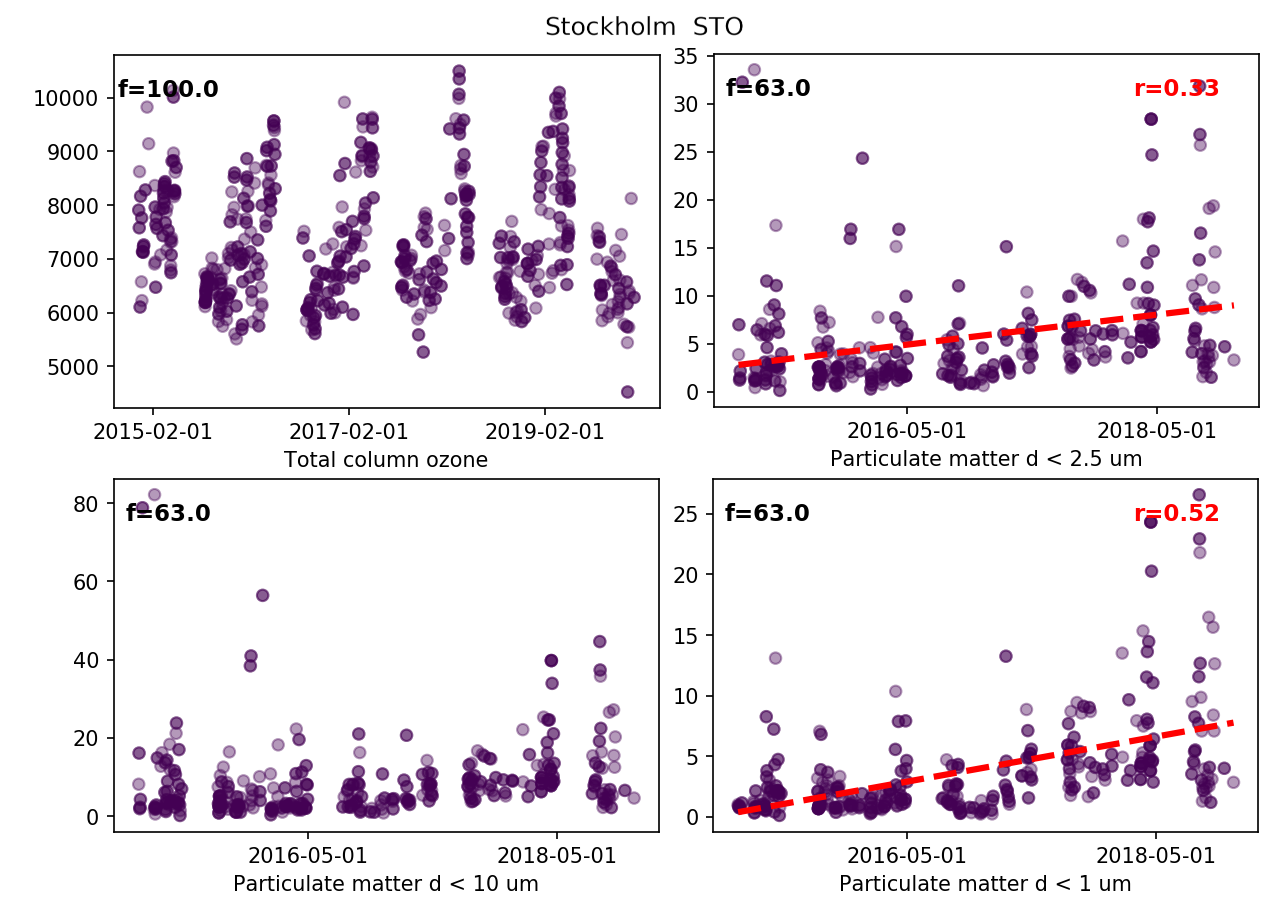}}
    
    \subfloat{\includegraphics[width=0.85\textwidth]{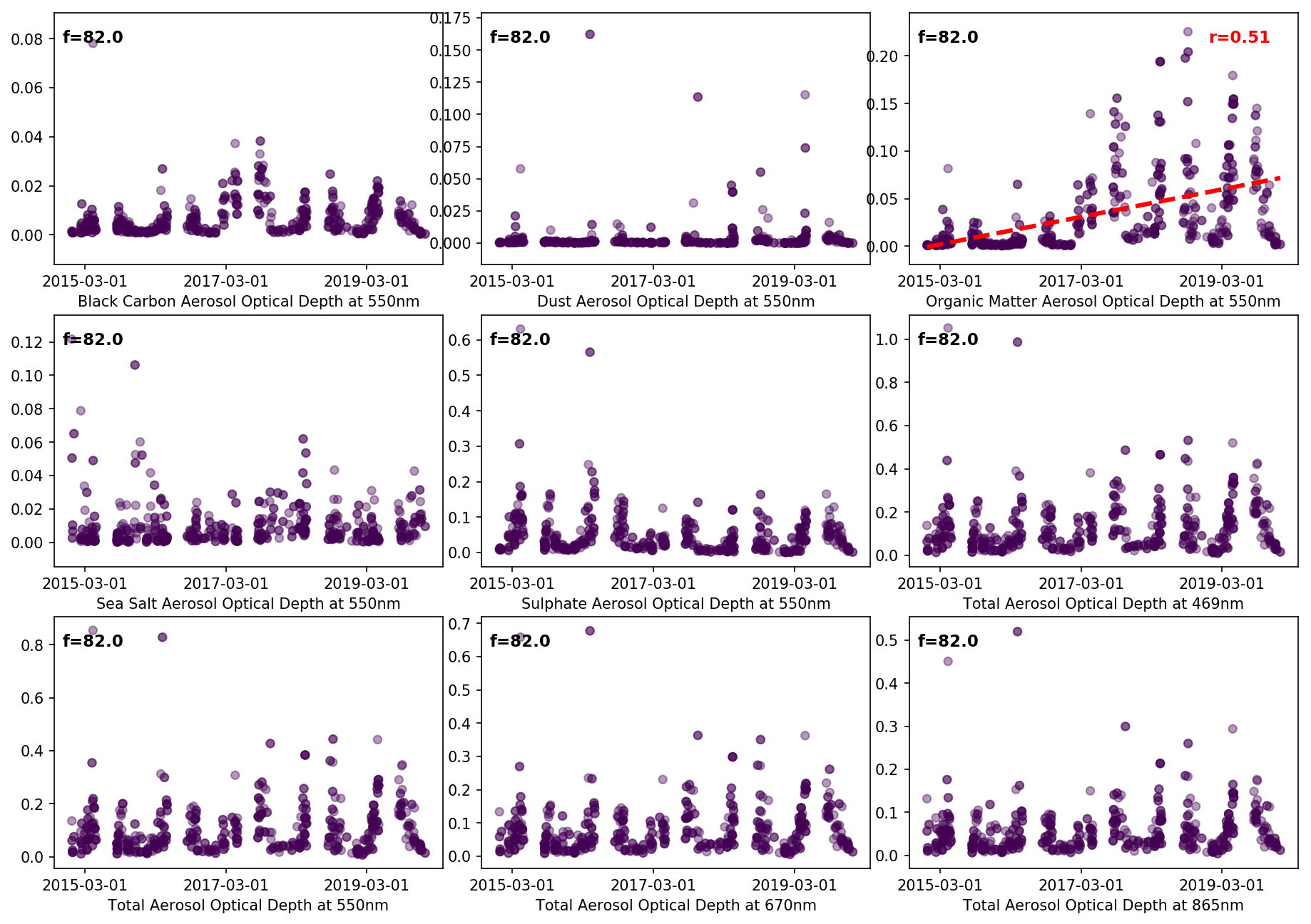}}
        
    \caption[Stockholm Meteo]{Large-scale particulate matter (\textit{top panel}) and aerosol optical depth (\textit{bottom panel}) development for Stockholm.
    The parameter `f' in the top left corner indicates what fraction of available SQM measurements could be matched with the meteorological data shown here.
    The y-axes units are $nm$ for molecular columns and $g\ cm^{-3}$ for particulate matter.}
    \label{fig:meteo3}
\end{figure*}

\bsp	
\label{lastpage}
\end{document}